\documentclass{mem}
\usepackage{natbib}\usepackage{txfonts}\usepackage{balance}
\usepackage{graphicx}
\usepackage[a4paper]{hyperref}
\idline{000}{1}

\def \sw {$Swift$}

\def \hcm {\hbox {\ifmmode $ atom cm$^{-2}\else atom cm$^{-2}$\fi}}

\def \ATel {ATel}

\defcitealias{SidoliPM2006}{Paper~I}
\begin{document}

\title{Supergiant Fast X--ray Transients: a Review}

  \author{P. Romano\inst{1} \and L.\ Sidoli\inst{2},  on behalf of the SFXT Gang  }
   \offprints{P.\ Romano}
   \institute{INAF, Istituto di Astrofisica Spaziale e Fisica Cosmica, 
	Via U.\ La Malfa 153, I-90146 Palermo, Italy 
        \email{romano@ifc.inaf.it}      
    \and INAF, Istituto di Astrofisica Spaziale e Fisica Cosmica, 
	Via E.\ Bassini 15,   I-20133 Milano,  Italy    
             }

\authorrunning{Romano \& Sidoli}

\titlerunning{Supergiant Fast X--ray Transients}

\abstract{We review the status of our knowledge on supergiant fast X--ray transients (SFXTs),
a new hot topic in multi wavelength studies of binaries.
We discuss the mechanisms believed to power these transients and then
highlight the unique contribution {\it Swift} is giving to this field, 
and how new technology complements and sometimes changes the view of things.

}
\maketitle{}

\section{Introduction}

The recognition of supergiant fast X-ray transients (SFXTs) as a new class 
of high-mass X--ray binaries is inevitably linked with the observations 
performed by {\it INTEGRAL} during its Galactic Plane survey, which started 
after the launch, in 2002. 

As testified by the IBIS/ISGRI catalogue \citep[][and references therein]{Bird2007}
about a third of the catalogued objects are cataclysmic variables and low-mass X--ray binaries
(CVs and LMXB are 13\,\% and 14\,\% of the total, respectively), 
38\,\% are active galactic nuclei (AGNs), and about 35\,\% are high-mass X--ray binaries (HMXBs).
Among the latter, several are intrinsically highly absorbed and were obviously more 
difficult to discover with previous missions \citep[e.g.\ IGR J16318--4848,][]{Walter2003}. 
Others are the newly recognized class of HMXBs, the supergiant fast X--ray transients (SFXT),
which constitute a new hot topic in multi-wavelength studies of binaries.

SFXTs are firmly associated with an O or B supergiant,
and display outbursts 
which are significantly shorter than typical Be/X-ray binaries and 
which are characterized by bright flares with a duration
of a few hours and peak luminosities of 10$^{36}$--10$^{37}$~erg~s$^{-1}$ 
\citep{Sguera2005,Negueruela2006}. 
The quiescence, characterized by a soft spectrum (likely thermal) and a 
low luminosity at $\sim 10^{32}$~erg~s$^{-1}$ is a rarely-observed state
(IGR~J17544--2619, \citealt{zand2005}; IGR J08408--4503, 
\citealt{Leyder2007}) 
which, combined with the outburst luminosity,  
yields a very large dynamic range of 3--5 orders of magnitude. 
Their hard X--ray spectra resemble the typical shape of HMXBs 
hosting X--ray pulsars, with a flat hard power law below 10\,keV, and a high 
energy cut-off at about 15--30~keV, sometimes strongly absorbed at soft energies 
\citep{Walter2006,SidoliPM2006}.
As their spectral properties are similar to those of accreting
pulsars, it is generally assumed that all members of the new class are 
HMXBs hosting a neutron star (NS), although the only SFXTs with a measured 
pulse period are
AX~J1841.0$-$0536 ($P_{\rm spin}\sim4.7$\,s, \citealt{Bamba2001}, 
IGR~J16465--4507 ($P_{\rm spin}\sim228$\,s, \citealt{Lutovinov2005}),  
IGR~J11215--5952 ($P_{\rm spin}\sim187$\,s, \citealt{Swank2007}), and   
IGR~J18483--0311 ($P_{\rm spin}\sim21$\,s, \citealt{Sguera2007}).  
Currently, there are 10 confirmed members of the SFXTs class 
and $\sim$20 more candidates which showed short transient flaring activity, 
but with no confirmed association with an OB supergiant companion.
The field is rapidly evolving, so this number is bound to increase in the near future.

\section{Outburst mechanisms}

The mechanisms responsible for the observed short outbursts are still being debated. 
The proposed explanations (see \citealt{Sidoli2009:cospar}, for a recent review) 
mainly involve the structure of the wind from the supergiant companion 
or the properties of the accreting NS.

\begin{enumerate}
\item {\bf Spherically symmetric clumpy winds}.  
In the spherically symmetric clumpy wind model, 
the short flares in SFXTs are supposed to be produced by accretion 
of massive clumps (10$^{22}$--10$^{23}$\,g) in the supergiant winds 
\citep{zand2005,Walter2007,Negueruela2008}, 
which are believed to be strongly inhomogeneous 
\citep[][]{Oskinova2007} with large density contrasts (10$^{4}$--10$^{5}$). 
In this model, the SFXTs should display wider orbits than persistent HMXBs.

\item  {\bf Equatorially enhanced wind.}  
Alternatively \citep{Sidoli2007}, the outbursts can be due to 
the presence of an equatorial wind component, denser, possibly clumpy, 
and slower than the symmetric polar wind from the blue supergiant, inclined with
respect to the orbital plane of the system. The enhanced accretion rate occurring when 
the NS crosses this wind component can explain SFXTs showing periodic outbursts, 
such as IGR~J11215--5952, as well as other SFXTs, by assuming different geometries of the 
equatorial wind. The recurrence in the outbursts from IGR J08408--4503 can indeed 
be explained within this model \citep{Romano2009:sfxts_paper08408}.

\item {\bf Gated mechanisms.} 
Other authors explain the high dynamic range in SFXTs with gated mechanisms 
\citep{Grebenev2007,Bozzo2008}  
where the accretion is halted by a magnetic
or a centrifugal barrier, dependent on the properties of the NS, 
its $P_{\rm spin}$ and its surface magnetic field $B$. 
In particular, \citet{Bozzo2008} conclude that SFXTs
should host neutron stars with long  $P_{\rm spin}\ga 1000$~s 
and magnetar-like $B\ga10^{14}$\,G fields.

\end{enumerate}

%%%%%%%%%%%%%%%%%%%%%%%%%%%%%%%%%%%%%%%%%%%%%%%%%%%%%%%%%%%%%%%%%%%%%%%% Table 
 \begin{table*}[t!]
 \begin{center}
 \caption{Summary of the {\it Swift}/XRT monitoring campaign.\label{vulcan09:tab:campaign} }
 \begin{tabular}{lcllll}
 \hline
 \noalign{\smallskip}
Name &Campaign & Exposure$^{\mathrm{b}}$ (N$^{\mathrm{a}}$)\hspace{-0.3cm} &Outburst$^{\mathrm{c}}$  &References \\
     &       Start         --End              &       & Dates      & \\
     &       ((yyyy-mm-dd)    & (ks)  & (yyyy-mm-dd)    & \\
  \noalign{\smallskip}
 \hline
 \noalign{\smallskip}
IGR~J16479$-$4514 &  2007-10-26--2008-10-25&  75.2 (70) & 2008-03-19       &\citet{Romano2008:sfxts_paperII}\\
                  &                        & 	        & 2008-05-21	       & \\ 
                  &   	                   & 	        & {\it 2009-01-29} & {\it \citet{Romano2009:atel1920}, } \\
                  &   	                   & 	        & &{\it \citet{LaParola2009:atel1929} } \\
XTE~J1739$-$302   &  2007-10-27--2008-10-31&  116.1 (95) & 2008-04-08       & \citet{Sidoli2009:sfxts_paperIII}\\
                  &   	                   & 	        & 2008-08-13     & \citet{Romano2008:atel1659}, \\
                  &   	                   & 	        & & \citet{Sidoli2009:sfxts_paperIV} \\
                  &   	                   & 	        & {\it 2009-03-10} & {\it \citet{Romano2009:atel1961} }\\ 
IGR~J17544$-$2619 &  2007-10-28--2008-10-31&  74.8 (77) & 2007-11-08		      & \citet{Krimm2007:ATel1265}\\
  		  &    	                   & 	        & 2008-03-31       & \citet{Sidoli2009:sfxts_paperIII}\\
  		  &   	        	   & 	        & 2008-09-04		       & \citet{Romano2008:atel1697},	\\
                  &   	        	   & 	        & 	  		       &\citet{Sidoli2009:sfxts_paperIV}   \\
                  &   	        	   & 	        & {\it 2009-03-15}	       &{\it \citet{Krimm2009:atel1971} }\\ 
AX~J1841.0$-$0536 &  2007-10-26--2008-11-15&  96.5 (88) & none	  		       &  \\  
  \noalign{\smallskip}
 \hline
 \noalign{\smallskip}
Total             &                        &  362.6 (330)                        & \\  
  \noalign{\smallskip}
  \hline
  \end{tabular}
  \end{center}
  \begin{list}{}{}
  \item[$^{\mathrm{a}}$]{Number of observations obtained during the monitoring campaign.}
  \item[$^{\mathrm{b}}$]{\sw/XRT net exposure.}
  \item[$^{\mathrm{c}}$]{BAT trigger dates. We report the outburst that occurred in 2009 in italics, for the sake of completeness. }
  \end{list}
  \end{table*}%%%%%%%%%%%%%%%%%%%%%%%%%%%%%%%%%%%%%%%%%%%%%%%%%%%%%%%%%%%%%%%%%%%%%%%% 

\section{Searching for the equatorial wind components in SFXTs with {\em Swift}}

Thanks to {\it Swift}, for the very first time, we have the chance to give SFXTs non serendipitous 
attention throughout all phases of their life, by observing them during the bright outbursts, 
the intermediate intensity state, and the quiescence. 
In this section we report on the results of an entire 
year of intense monitoring campaign with {\it Swift} of a sub-sample of 4 SFXTs, 
IGR J16479--4514, XTE J1739--302, IGR J17544--2619, and AX J1841.0--0536, which  
were chosen as confirmed SFXTs, i.e.\ they display both a `short' transient (and recurrent) 
X--ray activity and they have been optically identified with
supergiant companions.  
XTE~J1739--302 and IGR~J17544$-$2619, are generally considered prototypical SFXTs
%: XTE~J1739--302 was the first transient which showed
%an unusual X--ray behavior \citep{Smith1998:17391-3021}, 
%recently optically associated with a blue supergiant \citep{Negueruela2006}.
AX~J1841.0$-$0536/IGR~J18410$-$0535, was chosen because at the time it was the only
SFXT, in addition to  IGR~J11215--5952, where a pulsar had been detected \citep{Bamba2001}.
Finally, IGR~J16479$-$4514 had displayed in the past a more frequent X--ray 
outburst occurrence than other SFXTs \citep[see, e.g.][]{Walter2007}, 
and offered an {\it a priori} better chance to be caught during an outburst.   

As part of our ongoing campaign, we obtained 2--3 observations ($\sim 1$\,ks long) 
per week per object. This observing pace naturally fits in the regular scheduling 
of $\gamma$-ray bursts, the main targets for {\it Swift}. 
Our goals were to fully characterize the long-term behavior of 
SFXTs, to determine the properties of their quiescent state (where the accumulation of
large observing time is needed to allow a meaningful spectral analysis of this 
faintest emission), to monitor the onset of the outbursts, and to measure the outburst
recurrence period(s) and duration.

During the first year (the program is still ongoing), 
we collected a total of 330 \sw\ observations, for a total net XRT 
exposure of $\sim 363$\,ks and distributed as shown in Table~\ref{vulcan09:tab:campaign}. 
The main results of this campaign can be found in 
\citet[][Paper I, X-ray out-of-bright-outburst emission]{Sidoli2008:sfxts_paperI}, 
\citet[][paper II, outburst of IGR J16479--4514]{Romano2008:sfxts_paperII}, 
\citet[][Paper III and IV, outbursts of XTE~J1739--302 and IGR~J17544$-$2619]{Sidoli2009:sfxts_paperIII,Sidoli2009:sfxts_paperIV},
and \citet[][Paper V, first year results]{Romano2009:sfxts_paperV}.

%%%%%%%%%%%%%%%%%%%%%%%%%%%%%%%%%%%%%%%%%%%%%%%%%%%%%%%%%%%%%%%%%%%%%%%% Figure 
\begin{figure*}[ht!]
\resizebox{\hsize}{!}{\includegraphics[clip=true]{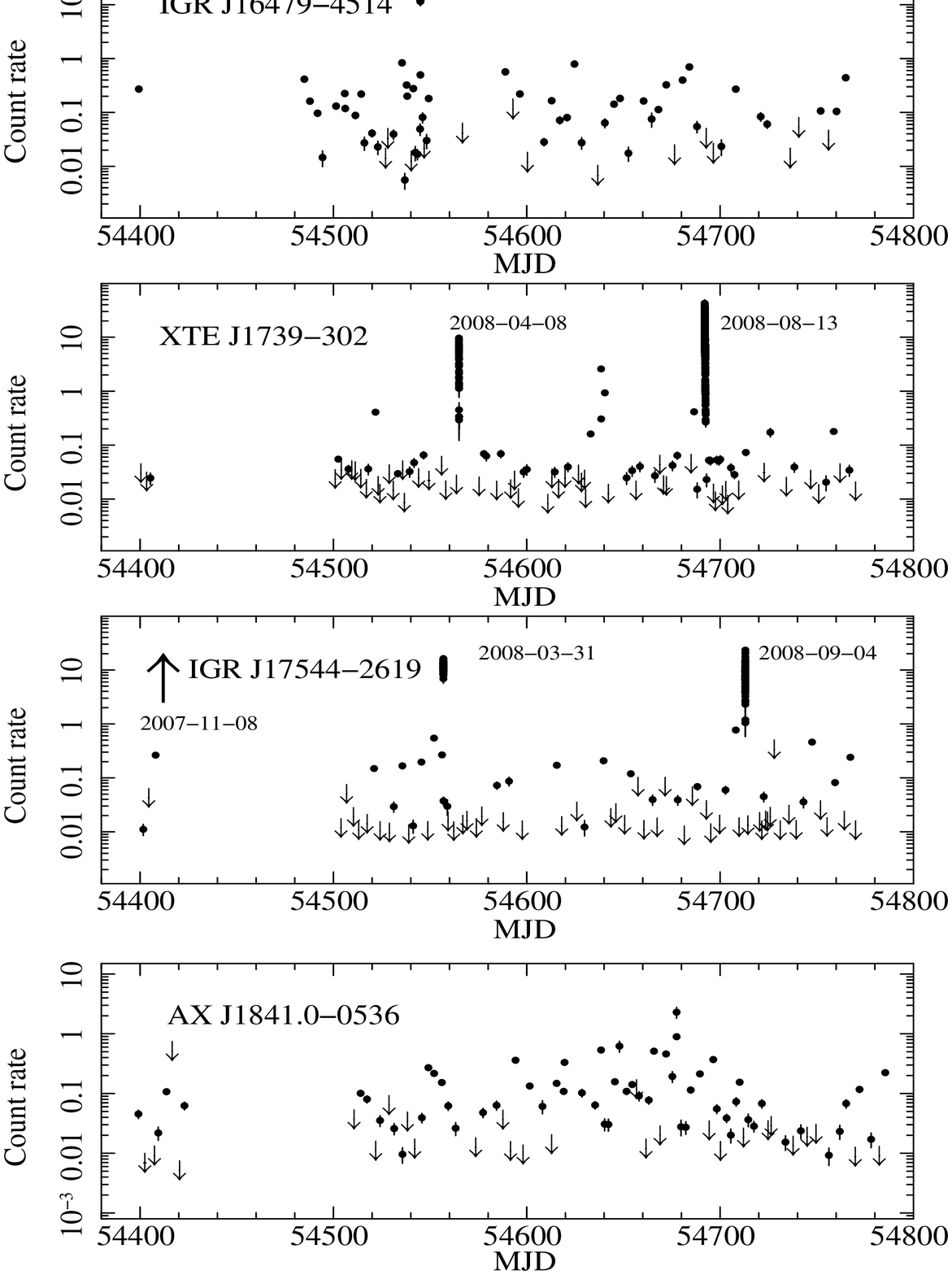}}
\vspace{-1truecm}
\caption{\footnotesize
{\it Swift}/XRT (0.2--10\,keV) light curves, corrected for pile-up, 
                PSF losses, vignetting and background-subtracted. The data were  
                collected from 2007 October 26 to  2008 November 15. 
		The downward-pointing arrows are 3-$\sigma$ upper limits. The upward pointing arrow 
                marks an outburst that triggered the BAT on MJD 54,414, but which could not be followed by XRT
                because the source was Sun-constrained for the XRT. Due to the sources being Sun-constrained
                between roughly 2007 December and 2008 January, depending on the target 
                coordinates, no data were collected during those months. 
                Adapted from \citet{Romano2009:sfxts_paperV}. 
}
\label{vulcan09:fig:xrtlcvs} 
\end{figure*}%%%%%%%%%%%%%%%%%%%%%%%%%%%%%%%%%%%%%%%%%%%%%%%%%%%%%%%%%%%%%%%%%%%%%%%% 

\subsection{XRT light curves}

% light curves
In Fig.~\ref{vulcan09:fig:xrtlcvs} we show the XRT 0.2--10\,keV light curves 
collected from 2007 October 26 to  2008 November 15. Each point in the
light curves refers to the average flux observed
during each observation performed with XRT; the exception are the outbursts 
(listed in Table~\ref{vulcan09:tab:campaign} and in Fig.~\ref{vulcan09:fig:xrtlcvs}) 
where the data were binned to include at least 20 source counts per time bin to best 
represent the count rate dynamical range.

\begin{table*}
 \begin{center}
 \caption{Spectroscopy of outbursts. \label{vulcan09:tab:xrtspectra}}
 \begin{tabular}{lcclllll}
 \hline
 \noalign{\smallskip}
Parameter &           IGR~J16479$-$4514 & IGR~J17544$-$2619 & XTE~J1739$-$302 \\
  \noalign{\smallskip}
 \hline
 \noalign{\smallskip}       
 Date & 2008-03-19 & 2008-03-31 &2008-04-08  \\   
$N_{\rm H}^{\mathrm{a}}$ ($10^{22}$~cm$^{-2}$)  & $6.2\pm0.5$ &       $1.1\pm0.2$  &    $12.5^{+1.5}_{-4.3}$ \\
$\Gamma$            & $1.15\pm0.14$   &   $0.75\pm0.1$  &   $1.4^{+0.5}_{-1.0}$\\
$E_{\rm cut}$  (keV)      & $6.6\pm0.9$ &	 $18\pm2$ &       $6^{+6}_{-7}$\\
$E_{\rm fold}$ (keV)      & $15.3\pm2.5$ &            $4\pm2$ &      $16^{+12}_{-8}$\\
$L_{\rm 0.5-10\,keV}^{\mathrm{b}}$ &   28  &     1.9 &   1.9  \\
$L_{\rm 0.5-100\,keV}^{\mathrm{b}}$ &  57  &    5.3 &      3\\
  \noalign{\smallskip}
Reference & \citet{Romano2008:sfxts_paperII}& \citet{Sidoli2009:sfxts_paperIII}&\citet{Sidoli2009:sfxts_paperIII} \\ 
  \noalign{\smallskip}
 \hline
 \noalign{\smallskip}
  \end{tabular}
  \end{center}
  \begin{list}{}{}
  \item[$^{\mathrm{b}}$]{In units of 10$^{36}$~erg~s$^{-1}$.}
 \end{list}
  \end{table*}%%%%%%%%%%%%%%%%%%%%%%%%%%%%%%%%%%%%%%%%%%%%%%%%%%%%%%%%%%%%%%%%%%%%%%%% 

\subsection{Outbursts}

\begin{figure*}
\resizebox{\hsize}{!}{\includegraphics[clip=true]{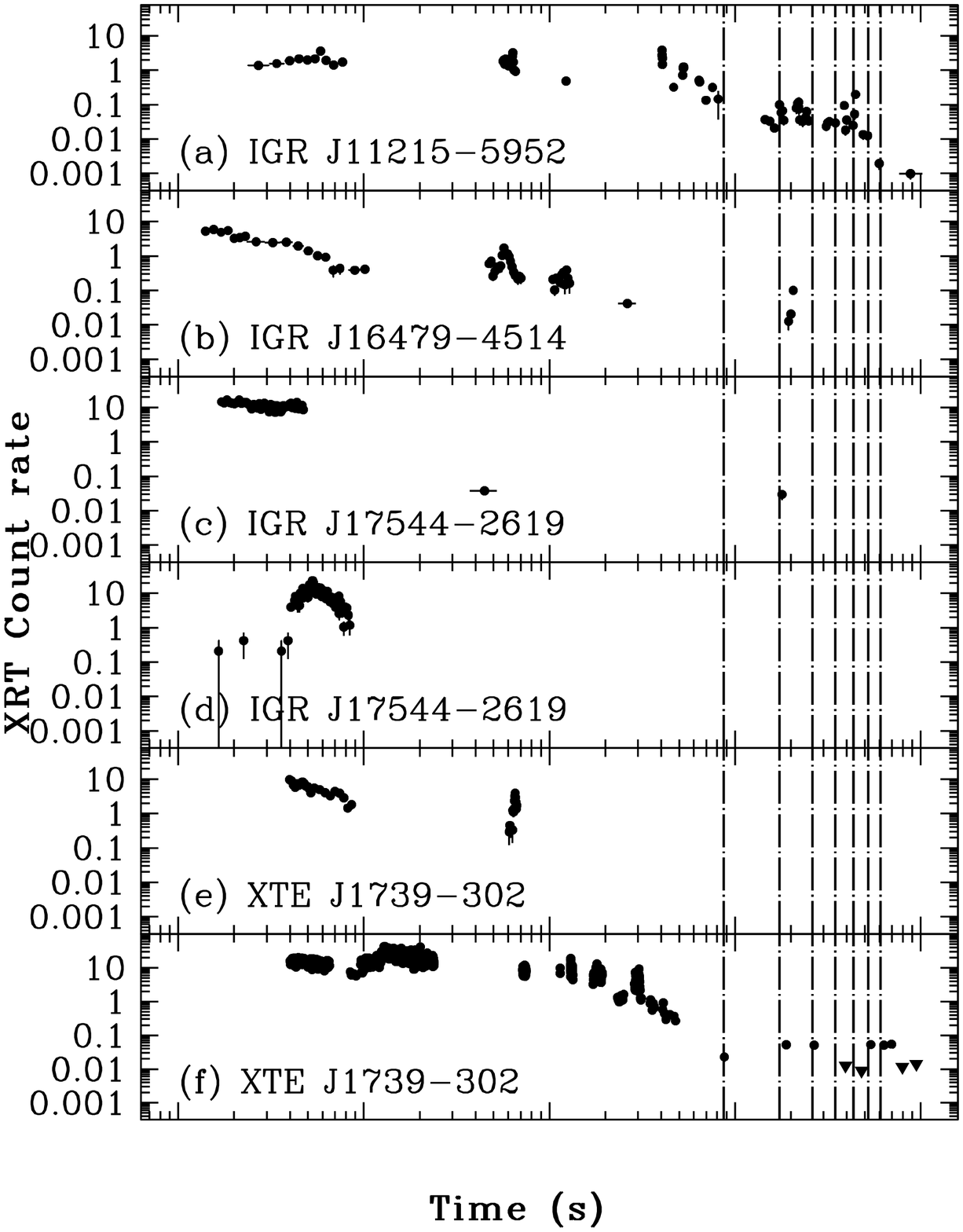}}
\caption{\footnotesize
{\it Swift}/XRT light curves of IGR~J11215--5952 (a), IGR~J16479$-$4514 (b), 
and of four of the outbursts observed 
during the observing campaign (see Table~\ref{vulcan09:tab:campaign}). 
}
\label{vulcan09:fig:xrtlcvsall} 
\end{figure*}

% OUTBURSTS (trigger and on board)
We obtained multi-wavelength observations of 5 outbursts of 3 different sources 
(see Table~\ref{vulcan09:tab:campaign}) during this first year of monitoring. 
AX~J1841.0$-$0536 is the only source which has not undergone a bright outburst, yet. 
The light curves are shown in Fig.~\ref{vulcan09:fig:xrtlcvsall}. 
As reported in 
\citet{Romano2008:sfxts_paperII}, \citet{Sidoli2009:sfxts_paperIV}, and 
\citet{Sidoli2009:sfxts_paperIII}, 
we examined the broad-band simultaneous spectra 
(0.3--150\,keV) of three SFXTs. They can be fit with absorbed 
cutoff power laws, which are models traditionally adopted for accreting X--ray 
pulsars even in the objects where proof of the presence 
of a neutron star (as derived from a spin period) is still unavailable. 
Considerable differences can be found in the
behaviour of the absorbing column among the examined cases.
In Table~\ref{vulcan09:tab:xrtspectra} we summarize the spectral 
fits of the first three outbursts we studied 
\citep{Romano2008:sfxts_paperII,Sidoli2009:sfxts_paperIV}.

Individual sources behave somewhat differently; nevertheless, 
common X--ray characteristics of this class are 
emerging such as outburst lengths well in excess of hours, with a
multiple peaked structure. We observed a high dynamic range (including bright outbursts) 
of $\sim4$ orders of magnitude in IGR~J17544$-$2619 and  XTE~J1739$-$302, 
of $\sim$3 in  IGR~J16479$-$4514, and of about 2 in 
AX~J1841.0$-$0536 (in the latter, due to the lack of bright flares).

\subsection{Inactivity duty cycle}

% INACTIVTY 
The light curves can be used not only to trace the activity states of these objects,
but also their inactivity, since they represent a casual sampling of the flux state
at an average resolution of $\sim 4$ days. Therefore, we can determine how long each source 
spends in each state using a systematic monitoring with a sensitive instrument. 

We define as {\it duty cycle of inactivity}, the time each source spends {\it undetected} 
down to a flux limit of (1--3)$\times10^{-12}$ erg cm$^{-2}$ s$^{-1}$ (the flux limit achieved
for an exposure of 900\,s) 
$${\rm IDC}= \Delta T_{\Sigma} / [\Delta T_{\rm tot} \, (1-P_{\rm short}) ] \, , $$  where  
$\Delta T_{\Sigma}$ is sum of the exposures accumulated in all observations, 
   each in excess of 900\,s, where only a 3-$\sigma$ upper limit was achieved, 
$\Delta T_{\rm tot}$ is the total exposure accumulated (Table~\ref{vulcan09:tab:campaign}, column 5), and 
$P_{\rm short}$ is the percentage of time lost to short observations (exposure $<900$\,s). 
We obtain that ${\rm IDC} \sim 17, 28, 39, 55$\,\%,  
for IGR~J16479$-$4514, AX~J1841.0$-$0536, XTE~J1739--302, and IGR~J17544$-$2619, respectively, with  
an estimated error of $\sim 5\,\%$ on these values. 
Therefore, compared with estimates from less sensitive instruments, 
true quiescence, which is below our detection ability even with the 
exposures we collected in one year, is a relatively rare state.
This demonstrates that these transients accrete matter throughout their lifetime at 
different rates.

\subsection{Out-of-outburst emission}

\begin{figure*}[t!]
\vspace{-2truecm}
\resizebox{\hsize}{!}{\includegraphics[clip=true]{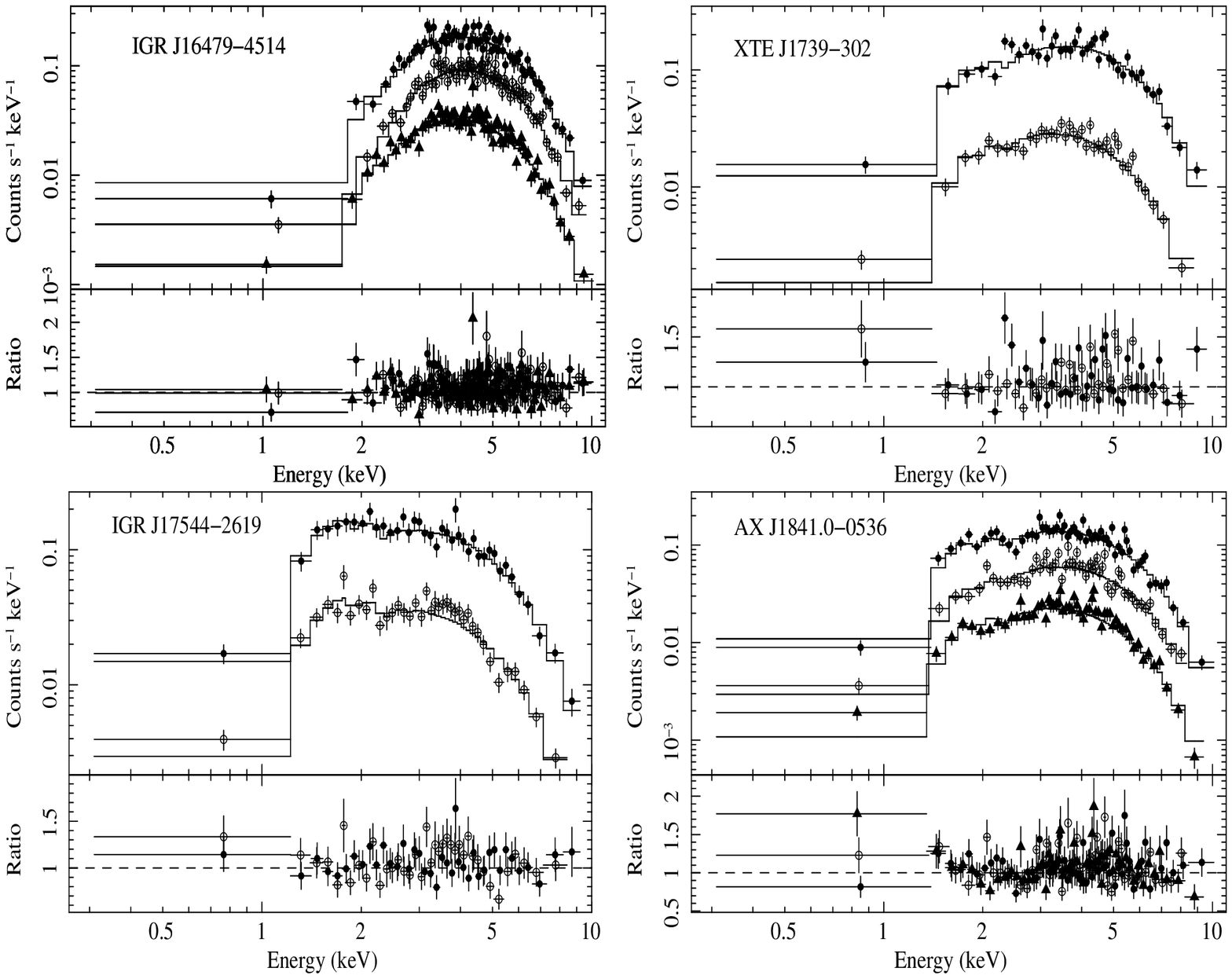}}
\vspace{-5truecm}
\caption{\footnotesize
Spectroscopy of the 2007-2008 {\it Swift} observing campaign. 
            Upper panels: XRT/PC data fit with an absorbed power law. 
            Lower panels: the data/model ratios. 
            Filled circles, empty circles, and filled triangles 
            mark high, medium, and low states, respectively. 
}
 \label{vulcan09:fig:ratios} 
\end{figure*}

% OUT-OF-OUTBURST emission 
Our {\it Swift}  monitoring campaign demonstrates for the first time that 
X--ray emission from SFXTs is still present outside the bright outbursts, 
although at a much lower level (10$^{33}$--10$^{34}$~erg~s$^{-1}$). 
This was already emerging from the first four months of this campaign 
\citep{Sidoli2008:sfxts_paperI}, 
but now we have accumulated enough statistics to allow intensity-selected 
spectroscopy of the out-of-outburst emission. 
When the spectra are fit with simple models (Fig.~\ref{vulcan09:fig:ratios}), 
such as an absorbed power law or a blackbody 
(more complex models were not required by the data), we obtain 
hard power law photon indices (always in the range $\Gamma$$\sim$0.8--2) or 
hot black bodies (kT$_{\rm BB}$$\sim$1--2~keV). 
In particular, when a blackbody model is adopted, the 
resulting radii of the emitter for all 4 SFXTs (and all the intensity 
states) are always only a few hundred meters. 
This clearly indicates an emitting region which is only a fraction of 
the neutron star surface, and can be naturally associated with the 
polar caps of the neutron star,  \citep{Hickox2004}. 
This evidence, coupled with the high level of flux variability and  
hard X--ray spectra, strongly supports the 
fact that the intermediate and low intensity level of SFXTs is 
produced by the accretion of matter onto the neutron star, demonstrating 
that SFXTs are sources which do not spend most of their 
lifetime in quiescence.

\section{Conclusions}

Let us review, as a conclusion, how each of the examined models 
for the outburst mechanisms fares when compared with the newly 
acquired {\it Swift} data. 

\begin{enumerate}
\item {\bf Clumping in spherically symmetric winds:} 
This model cannot explain the outburst periodicity
and the light curve shape (width) of IGR~J11215--5952. 
However, it could explain the low level of accretion
at $10^{33}-10^{34}$ erg s$^{-1}$ outside the bright outburst
assuming a distribution in the clump sizes or masses. 

\item {\bf Gated mechanism: SFXTs as magnetars:}

These models can reproduce the {\it Chandra} observation 
of IGR~J17544--2619 \citep{zand2005} 
and need a variable wind mass loss rate. 
However, they also require both long spin periods
(much shorter than the spin periods have been
measured until now in 4 SFXTs) and 
magnetar magnetic fields ($10^{14}-10^{15}$ G, 
but cut-off power laws observed in our sample
of SFXTs are compatible with low and more
standard surface NS magnetic fields). 
Furthermore, 
they cannot explain on their own the periodic IGR~J11215--5952.

\item {\bf Equatorially enhanced wind:}
This model explains the outburst periodicity and light curve shape 
observed in IGR~J11215--5952 \citep{Romano2007,Sidoli2007,Romano2009:11215_2008}.
It also naturally explains the low level of accretion outside the bright outbursts 
 as accretion from the polar wind component. 
Its key prediction is that the outburst recurrence should be periodic or almost periodic 
(or with a double/triple periodicity); alternatively, outbursts should be at least phase-locked, 
or concentrate preferentially in a particular orbital phase. 
Periodicities are starting to be found: we have indications of 
outbursts occurring every 11 and 24 days in IGR~J08408--4503  \citep{Romano2009:sfxts_paper08408};
a quasi-periodicity at $\sim 150$ days was found in IGR~J17544--2619 \citep{Sidoli2009:sfxts_paperIII}; 
finally, IGR~J18483--0311 shows a periodicity of 18.5 days \citep[based on INTEGRAL data][]{Sguera2007}.

For all other sources a longer monitoring is required since the expected 
orbital periods are in the order of months, and the picture could be much 
more complicated than in IGR~J11215--5952.

\end{enumerate}

\vspace{2truecm}

%%%%%%%%%%%%%%%%%%%%%%%%%%%%%%%%%%%%%%%%%%%%%%%%%%%%%%%%%%%%%%%%%%%%%%%% Figure 
%%%%%%%%%%%%%%%%%%%%%%%%%%%%%%%%%%%%%%%%%%%%%%%%%%%%%%%%%%%%%%%%%%%%%%%% 

%%%%%%%%%%%%%%%%%%%%%%%%%%%%%%%%%%%%%%%%%%%%%%%%%%%%%%%%%%%%%%%%%%%%%%%% Figure 
%%%%%%%%%%%%%%%%%%%%%%%%%%%%%%%%%%%%%%%%%%%%%%%%%%%%%%%%%%%%%%%%%%%%%%%% 

\begin{acknowledgements}
We would like to thank the organizers of this exceptional meeting and Franco, 
in particular, for running this workshop so very smoothly. 
This large project would have never flown without the contributions of 
our many collaborators, the SFXT Gang (G.\ Cusumano, V.\ La Parola, 
J.A.\ Kennea and S.\ Vercellone) first in line. 
We would like to thank the whole {\it Swift} Team starting from 
D.N.\ Burrows, S.\ Barthelmy, H.K.\ Krimm, J.A.\ Nousek, and N.\ Gehrels 
for their continued help and support, 
the science planners and duty scientists, for their professional juggling 
of a multitude of conflicting requests.  
This work was supported by contracts ASI I/088/06/0 and I/023/05/0. 
\end{acknowledgements}

\bigskip 
\bigskip 
\noindent {\bf DISCUSSION} 

\bigskip 
\noindent {\bf JOERN WILMS:}  
I would like to throw my weight behind the clumpy wind model. 
The reason is that we have seen behavior similar to that in SFXTs in HMXB. 
A key object is Vela X-1, where Staubert et al. and Kreykenbohm et al. 
have seen flares that look similar to SFXT flares. A clumpy wind model with an 
equatorial enhancement similar to that you suggest would be a good explanation for
Vela X-1 as well, i.e., we'd get a sequence HMXB--Vela X-1-like sources--SFXTs.

\bigskip 
\noindent {\bf P. ROMANO:} 
I'm glad you mentioned this. 
Indeed, there is a nice paper by Lorenzo Ducci \citep{Ducci2009}, 
who developed a stellar wind model for OB supergiants with a distribution for the 
masses and initial dimensions of the clumps. This model, together with the Bondi--Hoyle theory 
of wind accretion modified to take into account the presence of clumps, allows us to compare with
the observed properties of both the light curves and luminosity distributions
of the flares in SGXBs and SFXTs. 
This model was successfully applied to three representative HMXBs:
two persistent supergiant systems (Vela X--1 and 4U 1700--377) and the Supergiant Fast X-ray
Transient IGR J11215--5952. For the latter source, we had to introduce
a denser equatorial component (still with a clumpy structure) in order to reproduce the 
flare duration.
So, I guess, we are moving in the same direction. 

\end{document}